\documentclass [12pt]{article}
\usepackage{graphicx}

\begin{document}

\begin{center}
{\large \bf Lithium Experiment on Solar Neutrinos to Weight the
CNO Cycle.}

\vskip 0.2in

\underline {A.Kopylov,} I.Orekhov, V.Petukhov, A.Solomatin\\
Institute of Nuclear Research of Russian Academy of Sciences \\
117312 Moscow,
Prospect of 60th Anniversary of October Revolution 7A, RUSSIA\\
M.Arnoldov\\ Institute of Physics and Power Engineering, Obninsk,
Bondarenko sq.1, RUSSIA
\end{center}

\begin{abstract}
The measurement of the flux of beryllium neutrinos with the
accuracy of about 10\% and CNO neutrinos with the accuracy 30\%
will enable to find the flux of pp-neutrinos in the source with
the accuracy better than 1\% using the luminosity constraint. The
future experiments on $\nu $e$^{-}$ scattering will enable to
measure with very good accuracy the flux of beryllium and
pp-neutrinos on the Earth. The ratio of the flux of pp-neutrinos
on the Earth and in the source will enable to find with very good
accuracy a mixing angle theta solar. Lithium detector has high
sensitivity to CNO neutrinos and can find the contribution of CNO
cycle to the energy generated in the Sun. This will be a stringent
test of the theory of stellar evolution and combined with other
experiments will provide a precise determination of the flux of
pp-neutrinos in the source and a mixing angle theta solar. The
work on the development of the technology of lithium experiment is
now in progress.
\end{abstract}

The remarkable progress achieved in a number of experiments with
solar neutrinos \cite{1} with a culmination of KamLAND \cite{2}
has shown unambiguously that solar neutrinos do oscillate and the
parameters of neutrino oscillations belong to the MSW LMA region
\cite{3}, which is now split into two sub-regions so that at
3$\sigma $ we have \footnote{The new results in SNO experiment
(SNO collaboration nucl-ex/0309004) have shown (at 1$\sigma$) that
$\Delta m^2 = 7.1^{+1.2}_{-0.6} eV^2$ and $\theta =
32.5^{+2.1}_{-2.3} degrees$}

\begin{center}
$5.1 \times 10^{-5} eV^2 < \Delta m^{2} < 9.7 \times 10^{-5}
eV^{2}$
\end{center}

\begin{center}
$1.2 \times 10^{-4} eV^{2} < \Delta m^{2} < 1.9 \times 10^{-4}
eV^{2}$
\end{center}

\noindent for a mixing angle $\theta _{\odot}$

\begin{center}
$0.29 < tan^2 \theta _{\odot} < 0.86$
\end{center}

The further progress can be achieved by increasing the accuracy of
measurements of neutrino fluxes. Here some new aspect has arisen
connected with the possibility to increase drastically the
accuracy in the evaluation of the contribution of pp chain to the
total luminosity of the Sun. The luminosity constrained was first
suggested on a quantitative basis by M.Spiro and D.Vignaud
\cite{4}. As it is known, there are basically two sources of solar
energy: the pp-chain of reactions and CNO-cycle, the latter is
presented on Fig.1. Each neutrino source contributes to the solar
luminosity a definite value, so that the energy balance can be
written:

\begin{center}
$0.913f_{pp} + 0.002f_{pep} + 0.07f_{Be} + 0.0071f_{N} +
0.0079f_{O} = 1$ \hskip 0.5in (1)
\end{center}

\noindent for all neutrino sources with the coefficient by the
reduced neutrino flux greater 0.0001. Here the neutrino fluxes $f_
x$ are given in a reduced (relative to the predicted ones by
standard solar model (SSM) BP2000 \cite{5}) and the solar
luminosity is normalized to 1. The coefficients by $f_{pp}$,
$f_{pep}$ and $f_{Be}$ were obtained from numbers of Table 1
presented in \cite{6}, the coefficients by $f_N$ and $f_O$ were
calculated by us accounting that the energy produced for each
$^{13}$N neutrino in a first half-cycle CNO

\begin{center}
$\alpha(^{13}N) = M(^{12}C) + 2M(^1H) - M(^{14}N) - \langle
E_{\nu} \rangle (^{13}N)$ \hskip 0.5in (2)\\ $\alpha(^{13}N)  =
11.00 MeV$
\end{center}

\noindent and for each $^{15}$O neutrino for the second half-cycle
CNO

\begin{center}
$\alpha(^{15}O) = M(^{14}N) + 2M(^1H) - M(^4He) - M(^{12}C) -
\langle E_{\nu} \rangle (^{15}O)$ \hskip 0.5in (3)\\
$\alpha(^{15}O) = 14.01 MeV$
\end{center}

\begin{figure}[!ht]
\centering
\includegraphics[width=3in]{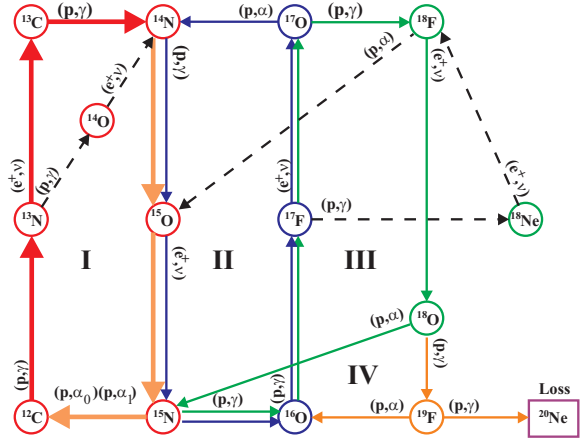}
\caption{The CNO cycle of thermonuclear reactions.}
\end{figure}

At the present epoch of the evolution of the Sun the CNO cycle is
closed, i.e. the concentration of $^{14}$N is close to the
equilibrium one as one can see from Fig.2 drawn by the numbers
presented in Table 4.4 of \cite{7}. On the early stage of solar
evolution the abundance of $^{14}$N in the interior of the Sun is
still low, consequently, the CNO cycle is not yet closed. From the
expression (1) it follows that beryllium neutrinos contribute to
the solar luminosity 7\% while CNO cycle (neutrinos from $^{13}$N
and $^{15}$O) -- only 1.5\% . Thus, according to SSM the
contribution of CNO cycle to the total energy generated in the Sun
is small but nevertheless, at a certain phase of experiment this
will be the principal limitation in evaluating the physical
quantities. For example, if the flux of beryllium neutrinos is
evaluated with the accuracy better than 10\% , the beryllium
uncertainty to the solar luminosity will be less than 1\% , so
that major uncertainty to the luminosity in this case will come
from CNO cycle. Then the measurement of the flux of neutrinos from
CNO cycle will the accuracy of about 30\% will enable to find the
flux of pp neutrinos in the source with the accuracy better than
1\% \cite{8}. At the present time a new generation of solar
neutrinos is under development. Some of these detectors are
oriented to measure precisely the flux of pp-neutrinos on the
Earth by means of $\nu $e$^{-}$ scattering \cite{9}. The
cross-section of this reaction is calculated with high accuracy so
principally they can do the high precision measurements. The
effect in $\nu $e$^{-}$ scattering experiment is determined mainly
by electron neutrinos, the contribution of neutrinos of other
flavors is small and well calculable. The ratio of these two
values: the flux of electron pp-neutrinos on the Earth and the
flux of pp-neutrinos in the source, i.e. in the Sun, will give the
survival probability and consequently, a mixing angle. This is a
very good approach. In some way it is similar to charged current
-- neutral current strategy in SNO experiment. But here one value
-- the flux of pp-neutrinos in the source - is found by measuring
the flux of beryllium and CNO neutrinos while other value -- the
flux of electron pp-neutrinos on the Earth is found by measuring
the effect in a $\nu $e$^{-}$ scattering detector. This strategy
becomes possible only because the contribution to the solar
luminosity of beryllium and CNO neutrino generated reactions is
relatively small, so that modest accuracy in the evaluation of the
flux of these neutrinos produces with high accuracy the flux of
pp-neutrinos in the source.

\begin{figure}[!ht]
\centering
\includegraphics[width=3in]{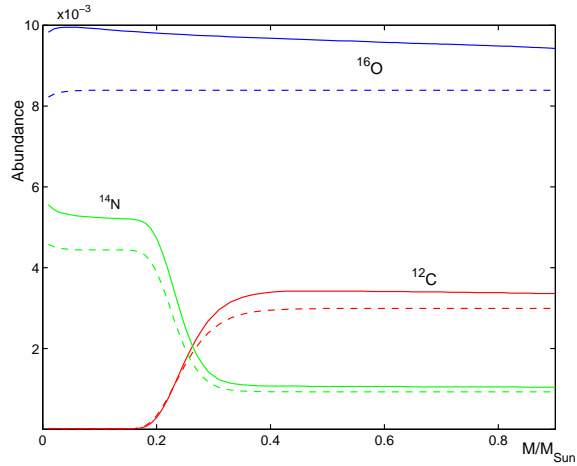}
\caption{The abundances of $^{12}C$ and $^{14}N$ in the interior
of the Sun. The solid lines - with diffusion in the matter of the
Sun, the dashed lines - without diffusion.}
\end{figure}

The present limit for the contribution of the CNO cycle is 7.3\%
\cite{10}. In a next few years even in the optimistic scenario of
the progress with the solar neutrino experiments it will be hardly
possible to decrease this limit lower than 5\% . So, future
experiments will not be able to exclude the weight of the CNO
cycle in the solar energy as much as 4.5\% . But in this case the
weight of a pp-chain may be not 98.5\% as is estimated now, but
only 95.5\% . The difference is 3\% and this is a principal
limitation. The accuracy in the evaluation of the flux of
pp-neutrinos generated in the Sun will be limited by this
uncertainty. So to go further one should measure the real weight
of CNO cycle. This can be done by a lithium detector because it
has high sensitivity to neutrinos generated in CNO cycle which
contribute 30\% to the expected rate. If the accuracy in a lithium
experiment is on the level of 10\% , then the weight of CNO
neutrinos will be determined with the uncertainty of about 0.5\%
in the absolute value of the contribution to the solar luminosity.
This result will be important for two reasons. First -- it will
provide a direct proof of the theory of stellar evolution. So far
we have no experimental evidence that CNO cycle does exist. These
data will show how correct is our understanding of the evolution
of the stars. For main sequence stars with higher temperature than
the Sun the CNO cycle is a major source of energy. In fact, the
fate of the Sun is also a CNO cycle star. Figure 2 shows the
expected concentrations of $^{12}$C and $^{14}$N in the interior
of the Sun.

One can see a peculiar distribution of these isotopes across the
radius of the Sun. Carbon is burned in nuclear reactions while
nitrogen is accumulated till the equilibrium concentration. To
test experimentally whether the real picture corresponds to our
understanding is very important. The second reason is that these
data combined with the results of other experiments will enable to
find with unprecedented accuracy the flux of pp neutrinos in the
source and from the ratio of the flux of electron neutrinos on
Earth to the one in the source - a mixing angle theta solar.

\begin{center}
\begin{table}[!ht]
Table 1. Standard Model Predictions (BP2000): solar neutrino
fluxes\\ and neutrino capture rates, with 1$\sigma$ uncertainties
from all sources\\ (combined quadratically).\\
\begin{tabular}
{|c|c|c|c|c|} \hline Source& Flux \par
(10$^{10}$cm$^{-2}$s$^{-1}$)& Cl
\par (SNU)& Ga \par (SNU)& Li \par (SNU) \\ \hline pp&
5.95(1.00$^{+0.01}_{-0.01}$)& 0.0& 69.7& 0.0
\\ \hline pep& 1.40$\times $10$^{-2}$(1.00$^{+0.015}_{-0.015}$)& 0.22& 2.8& 9.2
\\ \hline hep& 9.3$\times $10$^{-7}$& 0.04& 0.1& 0.1
\\ \hline $^{7}$Be& 4.77$\times $10$^{-1}$(1.00$^{+0.10}_{-0.10}$)& 1.15& 34.2& 9.1
\\ \hline $^{8}$B& 5.05$\times $10$^{-4}$(1.00$^{+0.20}_{-0.16}$)& 5.76& 12.1& 19.7
\\ \hline $^{13}$N& 5.48$\times $10$^{-2}$(1.00$^{+0.21}_{-0.17}$)& 0.09& 3.4& 2.3
\\ \hline $^{15}$O& 4.80$\times $10$^{-2}$(1.00$^{+0.25}_{-0.19}$)& 0.33& 5.5& 11.8
\\ \hline $^{17}$F& 5.63$\times $10$^{-4}$(1.00$^{+0.25}_{-0.25}$)& 0.0& 0.1& 0.1
\\ \hline Total& & 7.6$^{+1.3}_{-1.1}$& 128$^{+9}_{-7}$& 52.3$^{+6.5}_{-6.0}$
\\ \hline
\end{tabular}
\end{table}
\end{center}

The rates in a lithium detector from different neutrino sources
are presented in Table1. If to take the mixing angle $tan^2
\theta$=0.41 the total rate will be 25 SNU, and the contribution
of CNO cycle is 30\% . To find the flux of CNO neutrinos one
should subtract the rates from pep, $^7$Be and $^8$B neutrinos.
Even the current accuracies in the evaluation of these fluxes and
of mixing angle theta solar enable to do it taking into
consideration that the accuracy of 30\% in the evaluation of the
fluxes from CNO cycle are adequate for the task as it was
explained earlier. The flux of pep-neutrinos is found from the
flux of pp-neutrinos because the ration of these fluxes is well
known. More accurate measurement of the flux of beryllium
neutrinos will help in a further correction of data. This will be
done in Borexino and KamLAND experiments. Thus for a further
progress in the study of solar neutrinos it is vital to measure
the capture rate by a lithium target. There is some ambiguity due
to the unknown factor: what is contribution of $^{13}$N and
$^{15}$O neutrinos? In fact, the situation in this aspect is quite
good for a lithium target because due to a relatively high
threshold of this detector the rate from $^{13}$N neutrinos is
much less than the one from $^{{\rm 1}{\rm 5}}$O neutrinos: the
ratio is approximately 1/5, see Table 1 taken from ref.6. This
helps in the interpretation of the data because the uncertainty of
the rate from CNO neutrinos depends mainly upon the uncertainty of
the $^{15}$O neutrino flux.

To find the contribution of beryllium neutrinos one should know
not only the flux of beryllium neutrinos but also the shape of the
energy spectrum of these neutrinos due to thermal broadening. The
details of this were discussed in \cite{11}. The point is that in
the laboratory conditions the $^7$Be line will not produce $^{{\rm
7}}$Be on lithium since the reaction of $^7$Be production is
reverse to electron capture by $^7$Be. If to consider electron
screening in terrestrial atoms, the energy of beryllium line is
even lower than a threshold for $^7$Be production. But in the Sun
high temperature produces the thermal broadening of the $^{{\rm
7}}$Be line, as it was discussed in \cite{12} and later was
computed with high accuracy by Bahcall \cite{13}. Because of this
some fraction of the line with the energy higher than the
threshold will produce $^7$Be. The effect is model dependent. The
fact that the measured flux of boron neutrinos is in a good
agreement with the one predicted by BP2000 shows that the model
gives the correct temperature map of the interior of the Sun,
hence there is good reasoning to believe that the thermal
broadening of the beryllium line is described by the model
correctly.

The substantial issue is that while the contribution of CNO cycle
to the solar energy is only 1.5\% , the weight of neutrinos from
CNO cycle in the production rate of $^7$Be on $^7$Li is about 30\%
, so that for the total capture rate expected for a lithium target
25 SNU, neutrinos from CNO cycle contribute 8 SNU. If we take the
parameters of neutrino oscillations from the allowed region, make
the estimates for a detector with 10 tons of lithium and take pure
statistical uncertainties, then the capture rate on a lithium
target can be measured with the accuracy of approximately 1 SNU
for 16 Runs total performed during 4 years of measurements. Here
the efficiency of counting of $^7$Be was taken about 90\% what is
principally possible to do by means of the cryogenic detectors
\cite{14}. After subtracting the rate from these three sources of
a pp-chain one gets the rate from neutrinos of CNO cycle. Now we
have two possibilities: first, we can take the ratio of $^{13}$N
to $^{15}$O neutrinos as a given one by a SSM, or we can find
separately the contribution of these two neutrino sources to the
total capture rate solving the system of two equations:

$$ \left\{
\begin{array}{rcl}
L_H+L_{CN}+L_{NO}&=&L_{\odot}\\ R_H+R_{CN}+R_{NO}&=&R_{Li}\\
\end{array}
\right. \hskip 0.5in (4)
 $$

Here $L_H$, $L_{CN}$ and $L_{NO}$ are the contributions to the
solar luminosity of pp-chain and two half-cycles of CNO cycle,
$R_{Li}$ , $R_H$ - the measured and estimated for the hydrogen
sequence rates in a lithium detector, $R_{CN}$ and $R_{NO}$ means
the rates from neutrinos born in $^{13}$N- and $^{15}$O-decays, R
= yL/4$\pi $r$_{SUN}^2 \varepsilon $, where r$_{SUN}$ -- the
distance from Sun to Earth, $\varepsilon $ is the energy
contributed to the Sun per one neutrino emitted in each half-cycle
of CNO-cycle and y -- the capture rate per one neutrino of
$^{13}$N- or $^{15}$O-spectra. One can see from these equations
that principally it is possible to find separately the fluxes of
$^{13}$N and $^{15}$O neutrinos. The contributions of the energies
associated with $^{13}$N and $^{15}$O neutrinos are close as one
can see from the expressions (2) and (3). What concerns the rate
of a lithium detector, the situation here is very different. The
contribution of the $^{15}$O-neutrino is 5 times bigger than the
one of $^{13}$N-neutrinos. In other words, the straight lines
corresponding to equations of the system (4) are not parallel and
the system of equations has a solution.

The idea to use a lithium target for the detection of solar
neutrinos was expressed on the eve of a solar neutrino research
\cite{15} and this subject was investigated in a number of papers
\cite{16}. Lithium experiment is a radiochemical experiment and as
a target metallic lithium is planned to be used. The solar
neutrinos are captured by $^7$Li (the abundance is 93\% ) and
$^7$Be is produced in a reaction:

\begin{center}
$^7Li + \nu  \to Be + e^{-}$
\end{center}

The isotope $^7$Be has a half life 53 days and is decayed to
$^7$Li by means of electron capture. The aim is to extract $^{{\rm
7}}$Be from lithium and to count the number of extracted atoms.
The measured capture rate is converted in a neutrino flux. The
main advantage of a lithium detector is that transition $^{{\rm
7}}$Be-$^7$Li is super allowed hence the cross-section can be
calculated with a very good accuracy what was done in \cite{13}.
How to extract beryllium from metallic lithium? The basic
principle is the following. The chemical compounds of beryllium
with nitrogen (Be$_3$N$_2$) and oxygen (BeO) have extremely low
solubility in lithium, on the level of 10$^{-13}$ mole percents.
At the beginning of the exposure some beryllium sample (about 10
mg) is introduced in the target (10 tons of lithium). This
concentration of beryllium is by far exceeding the equilibrium
quantity. Beryllium is capturing nitrogen or oxygen atoms which
are always present in lithium as impurities and is converted in a
beryllium nitride or beryllium oxide. The same happens with the
beryllium atom produced by solar neutrinos. These compounds of
beryllium are ready to precipitate on any surface provided they
have the contact with it. This condition is rarely realized in a
bulk of lithium because the ratio of the surface to the volume is
negligible. But if the stream of lithium is passed through a fine
mesh filter with a very large surface, this contact is guaranteed.
So the extraction procedure consists in pumping lithium through a
filter and then by collecting the beryllium atoms by means of
extraction from an aqueous phase where all beryllium present on a
lithium film from the filter gets dissolved. The optimization of
the technology consists in finding the conditions by which the
extraction process is efficient. Initially the work was done with
the samples of $^7$Be produced in lithium by protons with the
energy of about 10-100 MeV. This is not very good technology
because $^{7}$Be is produced mainly on the surface of the sample,
not inside of a lithium samples and the extraction of $^7$Be
produced by this method was far from the real experimental
conditions. Now we use 14 MeV neutrons of D-T neutron source as a
$^7$Be generator. By this method one can guarantee that $^7$Be was
really produced inside of a lithium sample. Another problem is how
to count the extracted atoms of $^7$Be? By the decay of $^7$Be the
Auger electron is released with the energy 55 eV. This energy is
too small for counting by the traditional counting technique.
Principally it is possible to use a cryogenic detector as it was
demonstrated in \cite{14}. This method is the only possibility to
get the high precision measurements because of high efficiency of
counting $^7$Be atoms. But as it was shown earlier, it is not
absolutely necessary to have the accuracy of measurements on the
level of 1 SNU. The good physical result will be obtained even
with the modest accuracy of about 2.5 SNU (10\% ). In this case
one can use the counting by a low background gamma spectrometer
with the efficiency of counting of only 8\%. Here the accuracy of
10\% will be achieved with 10 tons of lithium and 4 years of
running the experiment. In 10\% $^7$Be decays to the excited level
of $^7$Li which emits the gamma with the energy 478 keV. This
energy is very convenient for counting. In the background spectrum
there is a very populated line 511 keV, so to discriminate the
background from this line one should use a gamma spectrometer with
a very good energy resolution. This can be realized by means of a
low background gamma spectrometer using high purity germanium
detectors \cite{17}. The assembly of the detectors which can be
used in this case is similar to one module planned to be used in a
Majorana project \cite{18}. The work on the development of the
technique of a lithium detector is now in progress at the
Institute of Nuclear Research RAS in Moscow, and in the Institute
of Physics and Power Engineering in Obninsk.

To summarize we should note that a lithium experiment is the only
way to find with unprecedented accuracy the flux of pp-neutrinos
generated in the interior of the Sun (in the source) by means of
measuring the fluxes of neutrinos generated in a CNO cycle. This
information is vital for further study of the thermonuclear
processes in the interior of the Sun and can be effectively used
for precise measurement of the mixing angle $\theta _{\odot}$. The
study of CNO is very important also as a precise test of the
theory of stellar evolution. This work was supported in part by
the Russian Fund of Basic Research, contract N 01-02-16167-A and
by the grant of Russia ``Leading Scientific Schools''
LSS-1782.2003.2. The authors deeply appreciate the very
stimulating discussions with G.Zatsepin, L.Bezrukov, V.Kuzmin,
V.Rubakov, S.Mikheev.


\begin{thebibliography}{99}

\bibitem{1} B.T.Cleveland et al., 1998, {\it Astrophys. J. \bf 496},
505\\ J.N.Abdurashitov et al., 2002, {\it J. Exp. Theor. Phys. \bf
95}, 181\\ W.Hampel et al. (GALLEX collaboration), 1999, {\it
Phys. Lett. \bf B447}, 127\\ T.Kirsten, 2002, {\it talk at the
XXth Int. Conf. On Neutrino Physics and Astrophysics (NU2002)},
Munich, May 25-30\\ M.Altmann et al., 2000, {\it Phys. Lett. \bf
B490}, 16\\ E.Bellotti et al. (GNO collaboration), 2000, in {\it
Neutrino 2000, Proc. of the XIXth Int. Conf. On Neutrino Physics
and Astrophysics},\\ 16-21 June, {\it eds. J.Law}\\ R.W.Ollerhead,
J.J.Simpson, 2001, {\it Nucl. Phys. B (Proc. Suppl.) \bf 91}, 44\\
Y.Fukuda et. al., 1996, {\it Phys. Rev. Lett. \bf 77}, 1638\\ S.
Fukuda et al., 2001, {\it Phys. Rev. Lett. \bf 86}, 5651\\
Q.R.Ahmad et al., 2001, {\it Phys. Rev. Lett. \bf 87}, 71301\\
Q.R.Ahmad et al., 2002, {\it Phys. Rev. Lett. \bf 89}, 11301

\bibitem{2} K.Eguchi et al. (KamLAND collaboration), {\it
hep-ex/0212021}

\bibitem{3} V.Barger, D.Marfatia, {\it hep-ph/0212126}\\
G.L.Fogli, E.Lisi, A.Marrone, D.Montanino, {\it hep-ph/0212127}\\
M.Maltoni, T.Schwetz, J.W.F.Valle, {\it hep-ph/0212129}\\
A.Bandyopadhyay, S.Choubey, R.Gandi, S.Goswami, {\it
hep-ph/0212146}\\ J.N.Bahcall, M.C.Gonzalez-Garchia,
C.Pe\~{n}a-Garay, {\it hep-ph/0212147}\\ P.C.de Hollanda and
A.Yu.Smirnov, {\it hep-ph/0212270}

\bibitem{4} M.Spiro and D.Vignaud, 1990, {\it Physics Letters B, \bf 242} 279-284

\bibitem{5} J.N.Bahcall, M.N.Pinsonneault, S.Basu, {\it
astro-ph/0010346}

\bibitem{6} J.N.Bahcall {\it astro-ph/0108148}

\bibitem{7} J.N.Bahcall, 1989, {\it Neutrino Astrophysics} Cambridge University Press,
{\it Cambridge}

\bibitem{8} A.Kopylov, 2003, {\it Lithium Project, Report at LowNU2003 Conference} Paris,
France\\ A.Kopylov and V.Petukhov, {\it hep-ph/0301016,
hep-ph/0306148}

\bibitem{9} M.Nakahata, 2003, {\it XMASS Project, Report at LowNU2003 Conference} Paris, France

\bibitem{10} J.N.Bahcall, M.C.Gonzalez-Garcia, and C.Pe\~{n}a-Garay, {\it
hep-ph/0212331}

\bibitem{11} J.N.Bahcall, 1994, The $^7$Be Solar Neutrino Line: A Reflection of the Central
Temperature Distribution of the Sun, {\it Preprint IASSNS-AST
93/40}

\bibitem{12} G.V.Domogatsky, 1969, {\it Preprint of Lebedev Phys. Inst., Moscow,
no.153}

\bibitem{13} J.N.Bahcall, Rev.Mod.Phys. 50 (1978) 881

\bibitem{14} M.Galeazzi, G.Gallinaro, F.Gatti et al., 1997,\\ {\it Physics Letters \bf B
398}, 187

\bibitem{15} J.N.Bahcall,1964, {\it Physics Letters \bf 3}, 332\\
G.T.Zatsepin, V.A.Kuzmin, 1965, {\it On the neutrino spectroscopy
of the Sun, in Proceedings of the 9th International Cosmic Ray
Conference, London}, 1024

\bibitem{16} J.K.Rowly, 1978, {\it Proc.Conf.on Status and Future of Solar Neutrino Research,
BNL}, Jan., 5-7, p.265\\
E.Veretenkin, V.Gavrin, E.Yanovich, 1985, {\it Russian Journal
``Atomic Energy'' \bf 88}, N1, 65\\
A.V.Kopylov, A.N.Likhovid, E.A.Yanovich, G.T.Zatsepin, 1993, {\it
Proc. Intern. School ``Particles and Cosmology'', Baksan Valley,
Russia, 22-27 April 1993. Editors: E.N.Alekseev, V.A.Matveev,
Kh.S.Nirov, V.A.Rubakov, World Scientific, Singapore-New
Jersey-London-Hong Kong}, p.63\\
S.Danshin, G.Zatsepin, A.Kopylov et al., 1997, {\it Part. Nucl.,
\bf 28}, 5\\
M.Galeazzi, G.Gallinaro, F.Gatti et al., 1997, {\it Physics
Letters \bf B 398}, 187\\
A.V.Kopylov, 2000, {\it in Proc.Intern.Conf. on Nonaccelerator New
Physics, Dubna, Russia, June 28 -- July 3, 1999, Russian Journal
``Nuclear Physics'', 2000, \bf 63, N7}, p.p.1345-1348

\bibitem{17} R.L.Brodzinski et al., 1990, {\it NIM \bf A292}, p.337

\bibitem{18} F.Avignone, 2003, {\it MAJORANA Project, Report at NANP2003 Conference}, Dubna, Russia

\end{thebibliography}
\end{document}